\documentclass[journal]{IEEEtran}
\usepackage{ifpdf}
\usepackage{cite}
\ifCLASSINFOpdf
   \usepackage[pdftex]{graphicx}
\else
   \usepackage[dvips]{graphicx}
\fi
\usepackage{epstopdf}
\usepackage[cmex10]{amsmath}
\usepackage{amsthm}

\usepackage{algorithmic}
\usepackage{array}
\ifCLASSOPTIONcompsoc
 \usepackage[caption=false,font=normalsize,labelfont=sf,textfont=sf]{subfig}
\else
 \usepackage[caption=false,font=footnotesize]{subfig}
\fi
 \usepackage{dblfloatfix}
\usepackage{url}
\begin{document}
%\bibliographystyle{IEEEtran}
\title{Weighted Proportional Fair Scheduling for Downlink Non-Orthogonal Multiple Access}

\author{Marie-Rita~Hojeij,~\IEEEmembership{Student~Member,~IEEE,}
        Charbel~Abdel Nour,~\IEEEmembership{Member,~IEEE,}
				Joumana~Farah,~\IEEEmembership{Member,~IEEE,}
        and~Catherine~Douillard,~\IEEEmembership{Senior~Member,~IEEE}% <-this % stops a space
\thanks{M. Hojeij, C. Abdel Nour and C. Douillard are with IMT-Alantique, CNRS UMR 6285 Lab-STICC, UBL, France (email: marie.hojeij@imt-atlantique.fr; charbel.abdelnour@imt-atlantique.fr; catherine.douillard@imt-atlantique.fr) and M. Hojeij is also with the Faculty of Engineering, 
Holy Spirit University of Kaslik, Lebanon}
\thanks{J. Farah is with the Faculty of Engineering, Lebanese University, Lebanon email:joumana.farah@ul.edu.lb}
\thanks{Part of this work has been performed in the framework of the Horizon 2020 project FANTASTIC-5G (ICT-671660), which is partly funded by the European Union. The authors would like to acknowledge the contributions of their colleagues in FANTASTIC-5G. This work has also been funded with support from the Lebanese University.}}
\maketitle

\begin{abstract}
In this paper, a weighted proportional fair (PF) scheduling method is proposed in the context of non-orthogonal multiple access (NOMA) with successive interference cancellation (SIC) at the receiver side. The new scheme introduces weights that adapt the classical PF metric to the NOMA scenario, improving performance indicators and enabling new services. The distinguishing value of the proposal resides in its ability to improve long term fairness and total system throughput while achieving a high level of fairness in every scheduling slot. 
Finally, it is shown that the additional complexity caused by the weight calculation has only a limited impact on the overall scheduler complexity while simulation results confirm the claimed improvements making the proposal an appealing alternative for resource allocation in a cellular downlink system. 
\end{abstract}

\begin{IEEEkeywords}
Non-orthogonal multiple access, scheduling, proportional fairness, resource allocation.
\end{IEEEkeywords}

% \fi
\IEEEpeerreviewmaketitle

\section{Introduction}
\label{sec:1}
\IEEEPARstart{R}{adio} access technologies apply multiple access schemes to provide the means for multiple users to access and share resources at the same time. In the 3.9 and fourth generation of mobile communication systems, such as Long-Term Evolution (LTE)\cite{ref4} and LTE-Advanced\cite{ref5,ref6}, orthogonal multiple access (OMA) based on orthogonal frequency division multiplexing (OFDM) or single carrier frequency division multiple access (SC-FDMA) were adopted, respectively for  downlink and uplink transmissions. Orthogonal multiple access techniques have gained their success from their ability to achieve good system-level throughput performance in packet-domain services, while requiring a reasonable complexity, especially due to the absence of multi-user detection.\\

However, with the proliferation of Internet applications, between the end of 2016 and 2022, total mobile traffic is expected to increase by 8 times\cite{refA}. At the same time, communications networks are required to further enhance system efficiency, latency, and user fairness.  To this end, non-orthogonal multiple access (NOMA) has recently emerged as a promising candidate for future radio access. By exploiting an additional multiplexing domain, the power domain, NOMA allows the cohabitation of two or more users per subcarrier. User multiplexing is conducted at the transmitter side, on top of the OFDM layer, and multi-user signal separation takes place at the receiver side, using successive interference cancellation (SIC)\cite{ref12,ref13,ref15,ref16,ref17,ref18,ref19}.\\

The main appeal of NOMA is that it improves user fairness while maximizing the total user throughput. The majority of existing works dealing with NOMA have investigated the system-level performance in terms of system capacity and cell-edge user throughput.\\
\indent In\cite{ref17}, NOMA using a SIC decoder is evaluated in comparison with OMA (i.e. when a subband is orthogonally divided in bandwidth and in power, among scheduled users). Simulation results show that system capacity and cell-edge user throughput are both increased.\\
\indent In\cite{ref23}, system-level performance, in terms of throughput, is assessed for an uplink non-orthogonal multiple access system. Optimized scheduling techniques are proposed and evaluated. A cost function is assigned to each possible pair of users, in order to maximize either the sum-rate or the weighted sum rate. The user pairing problem is solved by the Hungarian method and significant improvements in sum rates and cell-edge rates are shown compared to OMA.\\
\indent In\cite{refPoor}, two different situations are investigated in order to evaluate the performance of a downlink NOMA system. In the first scenario, the outage probability is considered as a performance evaluation metric, and each user has a target data rate based on its QoS. In the second scenario, the ergodic sum rate achieved by NOMA is assessed, and resources are allocated opportunistically according to encountered channel states, with no constraint on target data rates. It is shown that, in both situations, if target data rates and total allocated power are adequately chosen, NOMA can achieve better performance than OMA.\\
\indent In\cite{M.R}, several new strategies for the allocation of radio resources (in terms of bandwidth and power) in a downlink NOMA system have been investigated and evaluated. The main objective of \cite{M.R} is to minimize the number of allocated subbands, while guaranteeing a requested service data rate for each user. In this sense, several design issues have been explored: choice of user pairing, subband assignment, optimal and suboptimal power allocation, dynamic switching to OMA. Simulation results show that the proposed resource allocation techniques provide better performance when NOMA is used, compared to OMA. \\
\indent In\cite{ref16}, the system frequency efficiency and user fairness of a NOMA system are investigated in comparison with OMA. For this purpose, universal frequency reuse and proportional fairness (PF) scheduler\cite{PF1,PF2} are adopted. A large user throughput gain is observed for users near the base station, but the gain achieved by cell-edge user throughput is shown to be rather limited.\\
\indent Aiming at further enhancing the gain of the cell-edge user, a weighted PF-based multiuser scheduling scheme is proposed in \cite{ref18} in the context of a non-orthogonal access downlink system. A frequency block access policy is proposed for cell-interior and cell-edge user groups using fractional frequency reuse (FFR), with significant improvements in the user fairness and system frequency efficiency. In \cite {ImprovedPF}, an improved downlink NOMA scheduling scheme based on the PF scheduler is proposed and evaluated. The proposed scheme aims at taking the fairness of the target frame into consideration. It shows improved performance compared to the conventional PF scheduler.\\
Similarly to the work done in \cite{ref18}, several papers have proposed weighted versions of the PF scheduler, with the aim of improving user fairness in the OMA context.\\
\indent In\cite{DesignOfFairWeights}, fair weights have been implemented for opportunistic scheduling of heterogeneous traffic types for OMA networks. For designing fair weights, the proposed scheduler takes into account the average channel status as well as resource requirements in terms of traffic types. Simulation analysis demonstrates the effectiveness of the proposed scheme in terms of resource utilization, and its flexibility with regards to network characteristics changes due to user mobility.\\
\indent In\cite{Compensated}, the problem of fairness deficiency encountered by the PF scheduler when the mobiles experience unequal pathloss is investigated. To mitigate this issue, a modified version of the PF scheduler introducing distance compensation factors has been proposed. This solution was shown to achieve both high capacity and high fairness.\\
\indent In \cite{AWeightedPF}, a weighted PF algorithm is proposed in order to maximize best-effort service utility. The reason behind introducing weight factors into the PF metric is to exploit the inherent near-far diversity given by the pathloss. The proposed algorithm enhances both best-effort service utility and throughput performance, with a complexity similar to the complexity of the conventional PF scheduler.\\
Combining the efficient resource allocation achieved by NOMA with an implementation of fair weights is the main contribution of this paper. We propose indeed a weighted PF metric where several designs of the introduced weights are evaluated. The proposed scheme aims at providing fairness among users for each channel realization. By doing so, not only short-term fairness is achieved but also user capacity and long-term fairness are enhanced accordingly. On the other hand, the proposed schemes mitigate the problem of zero-rate incidence, inherent to PF scheduling, by attempting to provide non-zero rate to each user in any time scale of interest. This will further enhance the quality of experience (QoE) of all users.\\

This paper is organized as follows: In Section II, we introduce the system model and give a general description of the NOMA-based PF scheduler. Section III details the proposed weighted schemes in the NOMA context. In Section IV, we apply the fair weights to a resource allocation system based on OMA. Simulation results are given and analyzed in Section V, while Section VI concludes the paper.

%\IEEEPARstart{T}{his} 
%I wish you the best of success.

 
%\hfill September 17, 2014

\section{System Description}
\label{sec:2}

\subsection{Basic NOMA System}
\label{sec:2.1}
In this section, we describe the basic concept of NOMA including user multiplexing at the transmitter of the base station (BS) and signal separation at the receiver of the user terminal.\\
\indent In this paper, a downlink system with a single input single output (SISO) antenna configuration is considered. The system consists of $K$ users per cell, with a total bandwidth $B$ divided into $S$ subbands.\\
Among the $K$ users, a subset of users $U_s=\{k_1, k_2, ..., k_n, ..., k_{n(s)}\}$, is selected to be scheduled over each frequency subband $s$, ($1\le s\le S$). The $n$th user ($1\le n\le n(s)$)  scheduled at subband $s$ is denoted by $k_n$, and $n(s)$ indicates the number of users non-orthogonally scheduled at subband $s$. At the BS transmitter side, the information sequence of each scheduled user at subband $s$ is independently coded and modulated resulting into symbol ${{x}_{s,{{k}_{n}}}}$ for the $n$th scheduled user. Therefore, the signal transmitted by the BS on subband $s$, $x_s$, represents the sum of the coded and modulated symbols of the $n(s)$ scheduled users: 
\begin{equation}
\label{Eq:1}
{{x}_{s}}=\sum\limits_{n=1}^{n(s)}{{{x}_{s,{{k}_{n}}}}},\; with \; \text{ E}\left[ {{\left| {{x}_{s,{{k}_{n}}}} \right|}^{2}} \right]={{P}_{s,{{k}_{n}}}}
\end{equation}
where ${{P}_{s,{{k}_{n}}}}$ is the power allocated to user $k_n$ at subband $s$. The received signal vector of user $k_n$ at subband $s$, ${{y}_{s,{{k}_{n}}}}$, is represented by:
\begin{equation}
\label{Eq:2}
{{y}_{s,{{k}_{n}}}}={{h}_{s,{{k}_{n}}}}{{x}_{s,{{k}_{n}}}}+{{w}_{s,{{k}_{n}}}}
\end{equation}
where ${{h}_{s,{{k}_{n}}}}$ is the channel coefficient between user $k_n$ and the BS, at subband $s$. ${{w}_{s,{{k}_{n}}}}$ represents the received Gaussian noise plus inter-cell interference experienced by user $k_n$ at subband $s$. Let ${P}_{max}$ be the maximum allowable power  transmitted by the BS. Hence, the sum power constraint is formulated as follows:
\begin{equation}
\label{Eq:3}
\sum\limits_{s=1}^{S}{\sum\limits_{n=1}^{n(s)}{{{P}_{s,{{k}_{n}}}}}}={{P}_{\max }}
\end{equation}
The SIC process \cite{ref25} is conducted at the receiver side, and the optimal order for user decoding is in the increasing order of the channel gains observed by users, normalized by the noise and inter-cell interference $h_{s,{{k}_{n}}}^{2}/{{n}_{s,{{k}_{n}}}}$, where ${{n}_{s,{{k}_{n}}}}$ is the average power of ${{w}_{s,{{k}_{n}}}}$. Therefore, any user can correctly decode the signals of other users whose decoding order comes before that user. In other words, user $k_n$ at subband $s$ can remove the inter-user interference from the $j$th user, $k_j$, at subband $s$, provided $h_{s,{{k}_{j}}}^{2}/{{n}_{s,{{k}_{j}}}}$ is lower than $h_{s,{{k}_{n}}}^{2}/{{n}_{s,{{k}_{n}}}}$, and it treats the received signals from other users with higher $h_{s,{{k}_{j}}}^{2}/{{n}_{s,{{k}_{j}}}}$ as noise \cite{ref13,ref26}.\\
Assuming successful decoding and no error propagation, and supposing that inter-cell interference is randomized such that it can be considered as white noise \cite{ref16,ref23}, the throughput of user $k_n$, at subband $s$, ${{R}_{s,{{k}_{n}}}}$, is given by:
\begin{equation}
\label{Eq:4}
{{R}_{s,{{k}_{n}}}}=\frac{B}{S}{{\log }_{2}}\left( 1+\frac{h_{s,{{k}_{n}}}^{2}{{P}_{s,{{k}_{n}}}}}{\sum\limits_{j\text{ }\in \text{ }{{\text{N}}_{s}},\text{ }\frac{h_{s,{{k}_{n}}}^{2}}{{{n}_{s,{{k}_{n}}}}}<\frac{h_{s,{{k}_{j}}}^{2}}{{{n}_{s,{{k}_{j}}}}}}{h_{s,{{k}_{n}}}^{2}{{P}_{s,{{k}_{j}}}}+{{n}_{s,{{k}_{n}}}}}} \right)
\end{equation}
It should be noted that most of the papers dealing with resource allocation in downlink NOMA \cite{ref17,ref20,ref24,ref26}, consider a maximum number of users per subband of two, in order to limit the SIC complexity in the mobile receiver, except for \cite{ref16} and \cite{ref27} where this number respectively reaches 3 and 4. However, in the last two cases, static power allocation is assumed, which simplifies the power allocation step but degrades throughput performance. It has also  been stated that the performance gain obtained with 3 or 4 users per subband is minor in comparison to the case with 2 users.

\subsection{Conventional PF Scheduling Scheme}
\label{sec:2.2}
The PF scheduling algorithm has been proposed to ensure balance between cell throughput and user fairness. Kelly et al.\cite{PF2} have defined the proportional fair allocation of rates, and used a utility function to represent the degree of satisfaction of allocated users. In \cite{dump}, the operation of the PF scheduler is detailed: at the beginning of every scheduling slot, each user provides the base station with its channel state (or equivalently its feasible rate). The scheduling algorithm keeps track of the average throughput $T_{k}(t)$ of each user in a past window of length $t_c$. In the scheduling slot $t$, user $k^*$ is selected to be served based on:
\begin{equation}
{{k}^{*}}=\arg \underset{k}{\mathop{\max }}\,\frac{{{R}_{k}(t)}}{{{T}_{k}(t)}}
\end{equation}
where $R_k(t)$ is the feasible rate of user $k$ for scheduling slot $t$.\\

In \cite{PF1}, an approximated version of the PF scheduler for multiple users transmission is presented. This version has been adopted in the majority of the works dealing with NOMA \cite{ref20,ref26,ref27} in order to select users to be non-orthogonally scheduled on available resources.\\
%The proportional fairness scheduler has been used extensively in OFDMA based systems in order to manage the assignment of radio resources between users. The choice of the PF scheduler is reasonable due to the balance that it provides between fairness and system capacity.\\
For a subband $s$ under consideration, the PF metric is estimated for each possible combination of users $U$, and the combination that maximizes the PF metric is denoted by $U_s$:
\begin{equation}
{{U}_{s}}=\underset{U}{\mathop{\arg \max }}\,\sum\limits_{k\in U}{\frac{{{R}_{s,k}}(t)}{{{T}_{k}}(t)}}
\end{equation}
${{R}_{s,k}}(t)$ denotes the instantaneous achievable throughput of user $k$ at subband $s$ and scheduling time slot $t$.\\
Note that the total number of combinations tested for each considered subband is:
\begin{equation}
{{N}_{U}}=\left( \begin{matrix}
   1  \\
   K  \\
\end{matrix} \right)+\left( \begin{matrix}
   2  \\
   K  \\
\end{matrix} \right)+...+\left( \begin{matrix}
   N(s)  \\
   K  \\
\end{matrix} \right)
\end{equation}
${{R}_{s,k}}(t)$ is calculated based on Eq. ~\ref{Eq:4}, whereas $T_{k}(t)$ is recursively updated as follows\cite{PF1} :  
\begin{equation}
\label{eq:6}
{{T}_{k}}(t+1)=\left( 1-\frac{1}{{{t}_{c}}} \right){{T}_{k}}(t)+\frac{1}{{{t}_{c}}}\sum\limits_{s=1}^{S}{{{R}_{s,k}}(t)}
\end{equation}
Parameter $t_c$ defines the throughput averaging time window. In other words, this is the time horizon in which we want to achieve fairness. $t_c$ is chosen to guarantee a good tradeoff between system performance (in terms of fairness) and system capacity. We assume in the following a $t_c$ window of 100 time slots. With a time slot duration equal to 1 ms, a 100 ms average user throughput $T_{k}(t)$ is therefore considered.\\

\section{Proposed Weighted NOMA-Based Proportional Fairness (WNOPF) Scheduler }
\label{sec:3}
The PF scheduler both aims at achieving high data rates and at ensuring fairness among users, but  it only considers long-term fairness. In other words, a duration of $t_c$ time slots is 
needed to achieve fairness among users. However, short-term fairness and fast convergence towards required performance is an important issue to be addressed in upcoming mobile standards\cite{refA}.\\
Since all possible combinations of candidate users are tested for each subband, a user might be selected more than once and attributed multiple subbands during the same time slot. On the other hand, it can also happen that a user will not be allocated any subband whenever its historical rate is high. Then, the user will not be assigned any transmission rate for multiple scheduling slots. This behavior can be very problematic in some applications, especially those requiring a quasi-constant QoE such as multimedia transmissions. In such cases, buffering may be needed. However, such a scenario may not be compatible with applications requiring low latency transmission.\\
Therefore, we propose several weighted PF metrics that aim at: 
\begin{itemize}
	\item enhancing the user capacity, thus increasing the total achieved user throughput;
	\item reducing the convergence time towards required fairness performance;
	\item enhancing fairness among users (both long-term and short-term fairness);
	\item limiting the fluctuations of user data rates;
	\item incorporating the delivery of different levels of quality of service (QoS).
\end{itemize}

The proposed scheduler consists of introducing fair weights into the conventional PF scheduling metric. The main goal of the weighted metrics is to ensure fairness among users in every scheduling slot.\\
To do so, we start by modifying the PF metric expression so as to take into account the status of the current assignment in time slot $t$. Therefore, the scheduling priority given for each user is not only based on its historical rate but also on its current total achieved rate (throughput achieved during the current scheduling slot $t$), as proposed in\cite{ImprovedPF}.\\
Scheduling is performed subband by subband and on a time slot basis. For each subband $s$, the conventional PF metric $PF_s^{NOMA}$ and a weight factor $W(U)$ are both calculated for each candidate user set $U$. Then, the scheduler selects the set of scheduled users $U_s$ that maximizes the weighted metric $PF_s^{NOMA}(U) \times W(U)$. The corresponding scheduling method is referred to as Weighted NOMA PF scheduler, denoted by $WPF^{NOMA}$. The resource allocation metric can be formulated as follows:
\begin{equation}
\begin{split}
%\begin{align}
  & WPF{_{s}^{NOMA}}(U)=PF{_{s}^{NOMA}}(U)\times W(U) \\ 
 & {{U}_{s}}=\underset{U}{\mathop{\arg \max }}\,WPF{_{s}^{NOMA}}(U) \\ 
%\end{align}
\end{split}
\end{equation}
Weight calculation for each candidate user set $U$ relies on the sum of the weights of the multiplexed users. 
\begin{equation}
W(U)=\sum\limits_{k\in U}{{{W}_{k}}(t)}
\end{equation}
with
\begin{equation}
{{W}_{k}}(t)=R_{avg}^{e}(t)-{{R}_{k}}(t),\text{  }k\in U
\end{equation}
$R_{avg}^{e}(t)$ is the expected achievable bound for the average user data rate in the current scheduling slot $t$. It is calculated as follows:
\begin{equation}
R_{avg}^{e}(t)=b.R_{avg}(t-1)
\end{equation}
Since we tend to enhance the achieved user rate in every slot, each user must target a higher rate compared to the rate previously achieved. Therefore, parameter $b$ is chosen to be greater than 1.\\
The average user data rate, $R_{avg}(t)$, used in (12), is updated at the end of each scheduling slot based on the following:
\begin{equation}
{{R}_{avg}}(t)=\frac{1}{K}\sum\limits_{k=1}^{K}{\sum\limits_{s=1}^{S}{{{R}_{s,k}}(t)}}
\end{equation}
where $R_{s,k}(t)$ is the data rate achieved by user $k$ on subband $s$.\\

On the other hand, $R_{k}(t)$, the actual achieved data rate by user $k$ during scheduling slot $t$, is calculated as:
\begin{equation}
{{R}_{k}}(t)=\sum\limits_{s\in {{S}_{k}}}{{{R}_{s,k}}(t)},\text{  }k\in U
\end{equation}
with $S_k$ the set of subbands allocated to user $k$ during time slot $t$. At the beginning of every scheduling slot, $S_k$ is emptied; each time user $k$ is being allocated a new subband, $S_k$ and $R_{k}(t)$ are both updated.\\

The main idea behind introducing weights is to minimize the rate gap among scheduled users in every scheduling slot, thus maximizing fairness among them. A user set $U$ is provided with a high priority among candidate user sets if it contains non-orthogonally multiplexed users experiencing a good channel quality on subband $s$, having low or moderate historical rates, or/and having large rate distances between their actual achieved rates and their expected achievable average user throughput. The highest level of fairness is achieved when all users reach the expected user average rate $R_{avg}^{e}(t)$. By applying the proposed scheduling procedure, we aim to enhance  long-term and  short-term fairness at the same time.\\

The scheduling metric $PF^{NOMA}$, defined in (6), can guarantee the proportional fairness criterion by maximizing the sum of users service utility which can be formally written as \cite{PF1}:\\
\begin{equation}
P{{F}^{NOMA}}=\underset{scheduler}{\mathop{\max }}\,\sum\limits_{k=1}^{K}{\log {{T}_{k}}}
\end{equation}

The proposed weighted metric WNOPF achieves higher service utility compared to the conventional PF scheduler, if:
\begin{equation}
\sum\limits_{k=1}^{K}{\log {{T}_{k}}}\ge \sum\limits_{k=1}^{K}{\log T_{k}^{'}}
\end{equation}
where the historical rates $T_{k}$ and $T_k\prime$ correspond to the schedulers using the WNOPF metric and the conventional PF metric, respectively.\\
\\
\textbf{Proposition 1}: To make (16) valid, for a NOMA-based system, the following inequality should be verified:
\begin{equation}
\prod\limits_{k=1}^{K}{E\left[ W\left( {{U}_{k}} \right)/\sum\limits_{U}{W\left( U \right)} \right]}\prod\limits_{k=1}^{K}{E\left[ {{R}_{s,k}} \right]}\ge \prod\limits_{k=1}^{K}{E\left[ R_{s,k}^{'} \right]}
\end{equation}
$E[R_{s,k}]$ and $E\left[ R_{s,k}^{'} \right]$ are the statistical average of the instantaneous transmittable rate of user $k$ on a subband $s$, when WNOPF and the conventional PF scheduler are applied respectively. $U_k$ denotes a scheduled user set containing user $k$, $U$ is a possible candidate user set, and $E\left[ W\left( {{U}_{k}} \right)/\sum\limits_{U}{W\left( U \right)} \right]$ is the statistical average of the normalized weight of the set $U_k$.
\begin{proof}

Equation (16) can be written as:\\
\begin{equation}
\prod\limits_{k=1}^{K}{{{T}_{k}}}\ge \prod\limits_{k=1}^{K}{T_{k}^{'}}
\end{equation}
If we consider that ${{T}_{k}}={{I}_{k,tot}}/\left( {t}_{c}\Delta T \right)$, where $I_{k,tot}$ is the total amount of information that can be received by user $k$, for a total observation time $t_c \Delta T$, and $\Delta T$ is the scheduling time slot length, we obtain:\\
\begin{equation}
\prod\limits_{k=1}^{K}{\frac{I_{k,tot}}{t_c\Delta T}}\ge \prod\limits_{k=1}^{K}{\frac{I_{k,tot}^{'}}{t_c\Delta T}}
\end{equation}
If we denote by $N_k$ the number of allocated time slots for user $k$ within $t_c$, and $n_k$ the statistical average of the number of allocated subbands to user $k$ per time slot, (19) can be re-written as:\\
\begin{equation}
\prod\limits_{k=1}^{K}{\frac{{N_k}{n_k}E\left[ {R_{s,k}} \right]\Delta T}{t_c\Delta T}}\ge \prod\limits_{k=1}^{K}{\frac{N_{k}^{'}n_{k}^{'}E\left[ R_{s,k}^{'} \right]\Delta T}{t_c\Delta T}}
\end{equation}
Using a simple rearrangement, we get:\\
\begin{equation}
\frac{\prod\limits_{k=1}^{K}{\left({N_k}/{t_c} \right)S\left( {n_k}/S \right)}}{\prod\limits_{k=1}^{K}{\left(N_{k}^{'}/{t_c} \right)S\left(n_{k}^{'}/S \right)}}\ge \frac{\prod\limits_{k=1}^{K}{E\left[ R_{s,k}^{'} \right]}}{\prod\limits_{k=1}^{K}{E\left[ {{R}_{s,k}} \right]}}
\end{equation}
If $Pr_k$ ($={N_k}/{t_c}$) denotes the probability of user $k$ being scheduled per time slot and $pr_k$ ($={n_k}/{S}$) the statistical average probability of user $k$ being scheduled per subband, (21) can be reformulated as:\\
\begin{equation}
\frac{\prod\limits_{k=1}^{K}{P{{r}_{k}}p{{r}_{k}}}}{\prod\limits_{k=1}^{K}{Pr_{k}^{'}pr_{k}^{'}}}\ge \frac{\prod\limits_{k=1}^{K}{E\left[ R_{s,k}^{'} \right]}}{\prod\limits_{k=1}^{K}{E\left[ {{R}_{s,k}} \right]}}
\end{equation}
$pr_k$ can be regarded as the statistical average probability of a set $U_k$ (= $E\left[ \Pr \left( {{U}_{k}} \right) \right])$, being chosen among all possible candidate sets $U$ to be scheduled per subband. It is calculated as follows:\\
\begin{equation}
p{{r}_{k}}=E\left[ \Pr ({{U}_{k}}) \right]=E\left[ \Pr \left( P{{F}^{NOMA}}\left( {{U}_{k}} \right)W\left( {{U}_{k}} \right) \right) \right]
\end{equation}
Since the conventional PF metric $PF^{NOMA}$ and the weight calculation are independent, $pr_k$ is equal to:\\
\begin{equation}
\begin{matrix}
   p{{r}_{k}} & =E\left[ \Pr \left( P{{F}^{NOMA}}\left( {{U}_{k}} \right) \right) \right]E\left[ \Pr \left( W\left( {{U}_{k}} \right) \right) \right]  \\
   {} & =pr_{k}^{'}E\left[ W\left( {{U}_{k}} \right)/\sum\limits_{U}{W\left( U \right)} \right]  \\
\end{matrix}
\end{equation}
where $E\left[ W\left( {{U}_{k}} \right)/\sum\limits_{U}{W\left( U \right)} \right]$ is the statistical average of the normalized weight of a set $U_K$. In other words, the higher the user's weight within a certain time slot, the more frequently it is scheduled on a subband.\\
Thus, we obtain:\\
\begin{equation}
\frac{\prod\limits_{k=1}^{K}{P{{r}_{k}}pr_{k}^{'}E\left[ W\left( {{U}_{k}} \right)/\sum\limits_{U}{W\left( U \right)} \right]}}{\prod\limits_{k=1}^{K}{Pr_{k}^{'}pr_{k}^{'}}}\ge \frac{\prod\limits_{k=1}^{K}{E\left[ R_{s,k}^{'} \right]}}{\prod\limits_{k=1}^{K}{E\left[ {{R}_{s,k}} \right]}}
\end{equation}
Note that, in a NOMA-based system, the probability of a user being scheduled per time slot remains the same when using the proposed weighted metric or the conventional PF metric, since  users are distributed with uniform and random probability over the entire network in each time slot. Thus, we adopt the following approximation:
\begin{equation}
P{{r}_{k}}\simeq Pr_{k}^{'}
\end{equation}
Additional observations and verifications related to this approximation are given in VII. Therefore, (25) and (26) can also be formulated as (17).
 
\end{proof}

Other configurations of rate-distance weights can also be introduced. A promising one is obtained by substituting (27) for (9) and (10):\\
\begin{equation}
{{U}_{s}}=\arg \underset{U}{\mathop{\max }}\,\sum\limits_{k\in U}{\frac{{{R}_{s,k}}(t)}{{{T}_{k}}(t)}{{W}_{k}}(t)},\text{  }k\in U
\end{equation}
Here, the conventional NOMA-based PF metric and the weights are jointly calculated for each user $k$ in candidate user  set $U$. By doing so, we assign to each user its weight while ignoring the cross effect $\frac{{{R}_{s,k|U}}(t)}{{{T}_{k|U}}(t)}{{W}_{k'|U}}(t)$ produced by (9), where $k$ and $k'$ are non-orthogonally multiplexed users in the same $U$. This joint-based incorporation of weights is denoted by J-WNOPF in the following evaluations.\\

\section{Proposed Weighted OMA-Based PF Scheduler (WOPF)}
In the majority of existing works dealing with fair scheduling, OMA-based systems are considered. For this reason, we propose to  apply the weighted proportional fair scheduling metric  introduced in this paper  to an OMA-based system as well. This  allows  the contribution of NOMA within our framework to be evaluated. In the OMA case, non-orthogonal cohabitation is not allowed. Instead, a subband $s$ is allocated to only one user, based on the following metric:
\begin{equation}
{{k}^{*}}=\arg \underset{k}{\mathop{\max }}\,\frac{{{R}_{s,k}(t)}}{{{T}_{k}(t)}}{{W}_{k}(t)}
\end{equation}
where $W_k(t)$ is the weight assigned to user $k$, calculated similarly to the weights in WNOPF.
The conventional OMA-based PF scheduling metric is denoted by $PF^{OMA}$, whereas the resulting scheduling algorithm combining  OMA with the proposed weighted PF is denoted by WOPF.\\

OMA can be regarded as a special case of NOMA where only one user is allowed to be scheduled per subband. Therefore, in order to achieve a higher user service utility with WOPF than with the conventional PF scheduler in OMA, Proposition 1, detailed and proven in Section III, should also be verified for an OMA-based system. For this purpose, (17) is modified as follows:\\
\begin{equation}
\prod\limits_{k=1}^{K}{E\left[ {{W}_{k}}/\sum\limits_{k}{{{W}_{k}}} \right]}\prod\limits_{k=1}^{K}{E\left[ {{R}_{s,k}} \right]}\ge \prod\limits_{k=1}^{K}{E\left[ R_{s,k}^{'} \right]}
\end{equation}
where $W_k$ is the weight assigned to user $k$.

Note that, as in the NOMA case, we assume that the probability of a user being scheduled per time slot remains the same when using the proposed weighted metric or the conventional PF metric.

\section{Proposed Scheduling Metric for the First Scheduling Slot}
In the first scheduling slot, the historical rates and the expected user average data rate are all set to zero. Hence, the selection of users by the scheduler is only based  on the instantaneous achievable throughputs. Therefore, fairness is not achieved in the first scheduling slot, and the following slots are penalized accordingly. To counteract this effect, we propose to treat the first scheduling slot differently, for all the proposed weighted metrics.\\ 
For each subband $s$, the proposed scheduling process selects $U_s$ among the candidate user sets based on the following criterion:
\begin{equation}
{{U}_{s}}=\arg \underset{U}{\mathop{\max }}\,\sum\limits_{k\in U}{\frac{{{R}_{s,k}}(t=1)}{{{R}_{k}}(t=1)}}
\end{equation}
Note that when WOPF is considered, the maximum number of users per set $U$ is limited to 1.\\
$R_{k}(t=1)$, the actual achieved throughput, is updated each time a subband is allocated to user $k$ during the first scheduling slot. By doing so, we give priority to the user experiencing a good channel quality with regard to its actual total achieved data rate, thus enhancing fairness in the first slot.

\section{Incorporation of Premium Services}
In this section, we propose some changes to the proposed weighted metrics in order to give the possibility of delivering different levels of suality of service. In other words, the proposed metrics should have the ability to provide different priorities to different users or to guarantee a certain level of performance to a data flow. To do so, (11) is modified as follows:
\begin{equation}
{{W}_{k}}(t)=R_{service}-{{R}_{k}}(t),\text{  }k\in U
\end{equation}
where $R_{service}$ is the data rate requested by a certain group of users, corresponding to a certain level of performance. As an example, we  detail an example of 3 services, although the proposed modifications can be applied to an arbitrary number of services. $R_{service}$  is then  defined as follows:
\begin{equation}
{{R}_{service}}=\left\{ \begin{matrix}
   {{R}_{basic}},\text{ if }k\text{ requests a basic service}  \\
   {{R}_{silver}},\text{  if }k\text{ requests a silver service}  \\
   {{R}_{gold}},\text{ if }k\text{ requests a gold service}  \\
\end{matrix} \right.\text{, }k\in U
\end{equation}
This modification aims to guarantee a minimum requested service data rate for each user and also tends  to enhance the overall achieved fairness between users belonging to the same group, i.e. asking for the same service.

\section{Numerical Results}
\label{sec:5}
\subsection{System Model Parameters and Performance Evaluation}
This subsection presents the system level simulation parameters used to evaluate the proposed scheduling techniques. The parameters considered in this work are based on existing LTE/LTE-Advanced specifications \cite{ref3GPP}. We consider a baseline SISO antenna configuration. The maximum transmission power of the base station is 46 dBm. The system bandwidth is 10 MHz and is divided into 128 subbands when not further specified. The noise power spectral density is $4.10^{-18}$ mW/Hz. Users are deployed randomly in the cell and the cell radius is set to 500 m. Distance-dependent path loss is considered with a decay factor of 3.76. Extended typical urban (ETU) channel model is assumed, with time-selectivity corresponding to a mobile velocity of 50 km/h, at the carrier frequency of 2 GHz. In both OMA and NOMA scenarios, equal repartition of power is considered among subbands, as considered in\cite{ref16,ref20,ref24}. In the case of NOMA, fractional transmit power allocation (FTPA)\cite{refSyst} is used to allocate power among scheduled users within a subband. Without loss of generality, NOMA results are shown for the case where the maximum number of scheduled users per subband is set to 2 ($n(s)=2$).\\
\indent As for  parameter $b$ in (12), after several testings, the best performance was observed for $b$ equal to 1.5. In fact, the system has a rate saturation bound with respect to parameter $b$, since when we further increase $b$, similar performance is maintained. 

\subsection{Performance Evaluation}
\indent In this part, we mainly consider four system-level performance indicators: achieved system capacity, long-term fairness, short-term fairness, and cell-edge user throughput.\\
Several techniques are evaluated and compared. The following acronyms are used to refer to the main studied methods:
\begin{itemize}
	\item $PF^{NOMA}$: conventional PF scheduling metric in a NOMA-based system;
	\item $WNOPF$: proposed weighted PF scheduling metric in a NOMA-based system;
	\item $J-WNOPF$: proposed Weighted PF scheduling metric with a joint incorporation of weights in a NOMA-based system;
	\item $PF_{modified}^{NOMA}$: a modified version of the PF scheduling metric proposed in\cite{ImprovedPF}, where the actual assignment of each frame is added to the historical rate;
	\item $PF^{OMA}$: conventional PF scheduling metric in an OMA-based system;
	\item $WOPF$: proposed weighted PF scheduling metric in an OMA-based system.
\end{itemize}

In order to assess the fairness performance achieved by the different techniques, a fairness metric needs to be defined first. Gini fairness index\cite{refGini} measures the degree of fairness that a resource allocation scheme can achieve. It is defined as:
\begin{equation}
G=\frac{1}{2{{K}^{2}}\overline{r}}\sum\limits_{x=1}^{K}{\sum\limits_{y=1}^{K}{\left| {{r}_{x}}-{{r}_{y}} \right|}}
\end{equation}
with\\
\begin{equation}
\overline{r}=\frac{\sum\limits_{k=1}^{K}{{{r}_{k}}}}{K}
\end{equation}
$r_k$ is the throughput achieved by user $k$. When long-term fairness is evaluated, $r_k$ is considered as the total  throughput achieved by user $k$ averaged over a time-window length $t_c$:
\begin{equation}
{{r}_{k}}=\frac{1}{{{t}_{c}}}\sum\limits_{t=1}^{{{t}_{c}}}{{{R}_{k}}(t)}
\end{equation}
Otherwise, when fairness among users is to be evaluated within each scheduling slot, short-term fairness is considered and $r_k$ is taken equal to $R_k(t)$, the actual  throughput achieved by user $k$ during scheduling slot $t$. \\
Gini fairness index takes values between 0 and 1, where $G=0$ corresponds to the maximum level of fairness among users, while a value of $G$ close to 1 indicates that the resource allocation scenario is highly unfair. \\

\begin{figure}[h]
\centering
\includegraphics[width=0.54\textwidth]{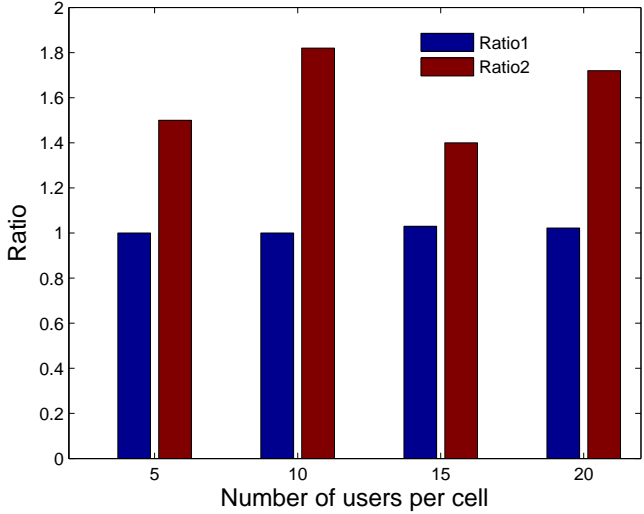}
\caption[Observed ratios related to (17) and (26) vs. number of users per cell.  NOMA-based system.]{Observed ratios related to (17) and (26) vs. number of users per cell, NOMA-based system.}
\end{figure}
First, we check the validity of Proposition 1 detailed in Section III and IV, and of the assumption done in (26). Fig. 1 shows the observed ratio between $Pr_k$ and $Pr_k\prime$, denoted by Ratio1, for different values of the number of users per cell. Fig. 1 also shows the ratio between the left hand and the right hand expressions of (17), denoted by Ratio2. Results show that Ratio1 is very close to 1, which means that the probability of a user being scheduled per time slot remains the same, under the proposed weighted metric or under the conventional PF metric. In addition, Ratio2 is shown to be greater than 1 regardless of the number of users per cell, which verifies Proposition 1, defined in (17). The results of a similar verification for an OMA system are observed in Fig. 2 .
\begin{figure}[ht]
\centering
\includegraphics[width=0.54\textwidth]{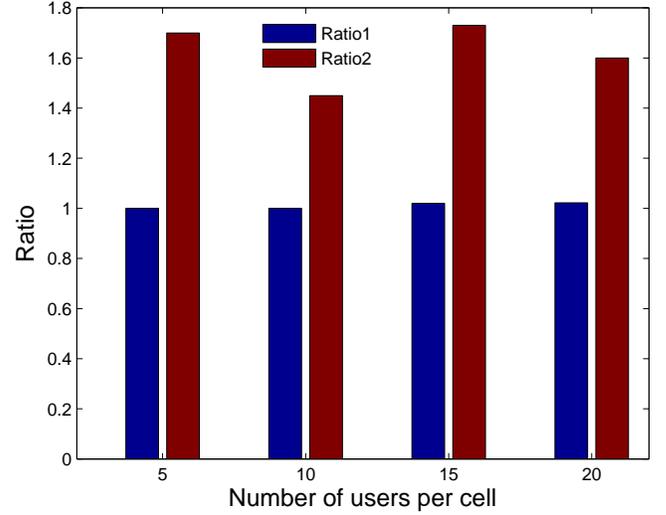}
\caption[Observed ratio related to (26) and (29) vs. number of users per cell. OMA-based system.]{Observed ratio   related to (26) and (29) vs. number of users per cell, OMA-based system.}
\end{figure}

\begin{figure}
\includegraphics[width=0.54\textwidth]{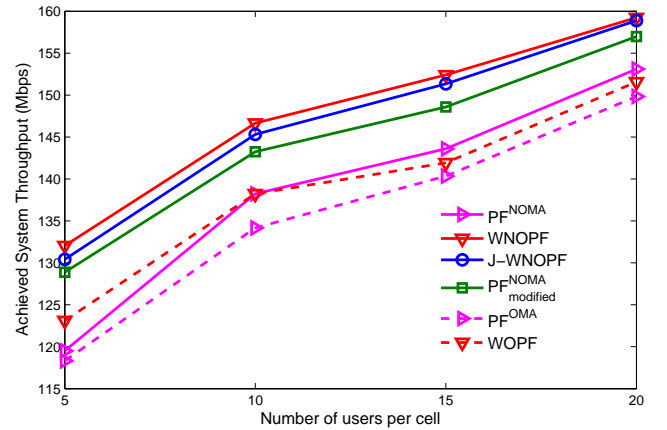}
% where an .eps filename suffix will be assumed under latex, 
% and a .pdf suffix will be assumed for pdflatex; or what has been declared
% via \DeclareGraphicsExtensions.
\caption{System throughput achieved with the proposed scheduling schemes vs. number of users per cell.}
\label{fig:1}
\end{figure}

Fig. 3 shows the system capacity achieved with each of the simulated methods for different numbers of users per cell. Curves in solid lines represent the NOMA case, whereas curves with dotted lines refer to OMA.\\ 
\indent We can observe that the throughput achieved with all the simulated methods increases as the number of users per cell is increased, even though the total number of used subbands is constant. This is due to the fact that the higher the number of users per cell, the better the multi-user diversity is exploited by the scheduling scheme, as also observed in\cite{DesignOfFairWeights}.\\
\indent The gain achieved by WNOPF, when compared to the other proposed weighted metric J-WNOPF, is mainly due to the fact that the joint incorporation of weights does not take into consideration the cross effect produced by non-orthogonally multiplexed users.\\
\indent The gain in performance obtained by the introduction of weights in the scheduling metric, compared to the conventional $PF^{NOMA}$ metric, stems from the fact that for every channel realization, the weighted metrics try to ensure similar rates to all users, even those experiencing bad channel conditions. With $PF^{NOMA}$, such users would not be chosen frequently, whereas appropriate weights give them a higher chance to be scheduled more often.\\
\indent Fig. 3 also shows an improved performance of the proposed metrics when compared to the modified PF scheduling metric $PF^{NOMA}_{modified}$ described in\cite{ImprovedPF}. Although they both consider the current assignment in their metric calculation, they still differ by the fact that the proposed weighted metrics target a higher rate compared to the rate previously achieved, therefore tending to increase the achieved user rate in every slot.\\
\indent When the proposed scheduling metrics are applied in an OMA context, WOPF provides higher throughputs than $PF^{OMA}$, due to the same reason why WNOPF outperforms $PF^{NOMA}$. Fig. 3 also shows a significant performance gain achieved by NOMA over OMA. All weighted scheduling metrics applying NOMA outperform the simulated metrics based on OMA, including WOPF. This gain is due to the efficient non-orthogonal multiplexing of users. It should also be noted that the gain achieved by WNOPF over $PF^{NOMA}$ is greater than the one achieved by WOPF over OPF: combining fair weights with NOMA definitely yields the best performance.\\

\begin{figure}
\centering
\includegraphics[width=0.52\textwidth]{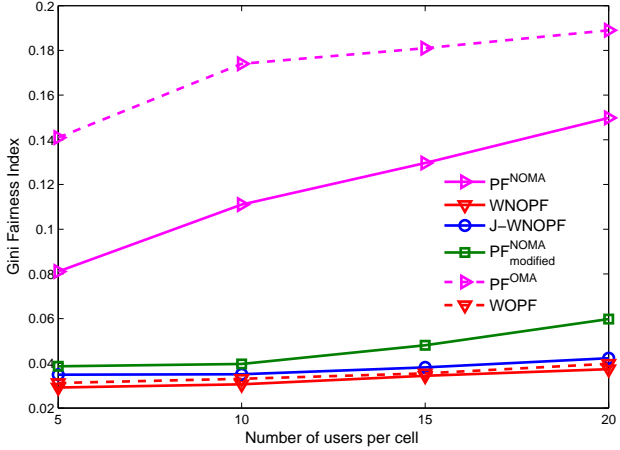}
% where an .eps filename suffix will be assumed under latex, 
% and a .pdf suffix will be assumed for pdflatex; or what has been declared
% via \DeclareGraphicsExtensions.
\caption{Gini fairness index of the proposed scheduling schemes vs. number of users per cell.}
\label{fig:2}
\end{figure}

Long-term fairness is an important performance indicator for the allocation process. Fig. 4 shows this metric as a function of the number of users per cell. Long-term fairness is improved when fair weights are introduced, independently of the access technique (OMA or NOMA). The reason is that, when aiming to enhance fairness in every scheduling slot, long-term fairness is enhanced accordingly. Again, in terms of fairness, the proposed weighted metrics outperform the modified PF metric\cite{ImprovedPF}, $PF_{modified}^{NOMA}$. This is due to the fact that WNOPF and J-WNOPF do not only consider the current rate assignment, but also tend to minimize the rate gap among scheduled users in every channel realization, thus maximizing fairness among them.\\

Fig. 5 shows the achieved system throughput as a function of the number of subbands $S$, for 15 users per cell. We can see that the proposed weighted metrics outperform the conventional PF scheduling scheme, for both access techniques OMA and NOMA, even when the number of subbands is limited. \\

\begin{figure}
\centering
\includegraphics[width=0.52\textwidth]{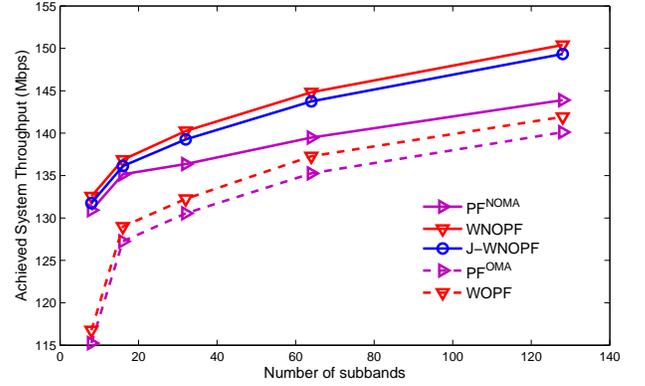}
% where an .eps filename suffix will be assumed under latex, 
% and a .pdf suffix will be assumed for pdflatex; or what has been declared
% via \DeclareGraphicsExtensions.
\caption{Achieved system throughput vs. $S$, for $K=15$.}
\end{figure}

\begin{figure}
\centering
\includegraphics[width=0.52\textwidth]{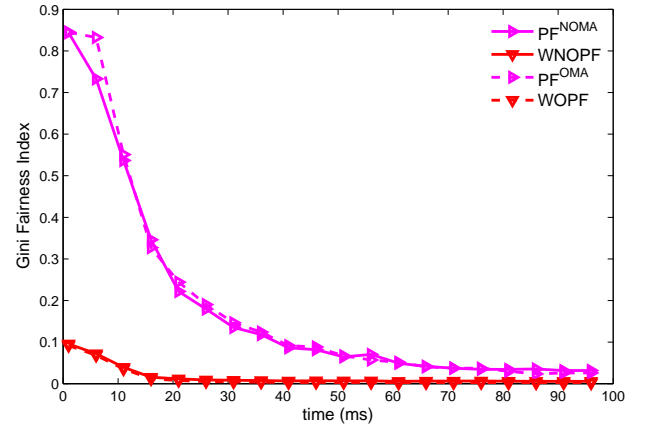}
% where an .eps filename suffix will be assumed under latex, 
% and a .pdf suffix will be assumed for pdflatex; or what has been declared
% via \DeclareGraphicsExtensions.
\caption{Gini fairness index vs. scheduling time index $t$.}
\end{figure}

Since WNOPF proves to give better performance than J-WNOPF, in terms of system capacity and fairness, J-WNOPF won't be considered in the subsequent results.\\
Since one of the main focuses of this study is to achieve short-term fairness, the proposed techniques should be compared based on the time required to achieve the final fairness level. Fig. 6 shows the Gini fairness index versus the scheduling time index $t$. The proposed weighted metric WNOPF achieves a high fairness from the beginning of the allocation process, and converges to the highest level of fairness (lowest value of index $G=0.0013$) in a limited number of allocation steps or time slots. On the contrary, $PF^{NOMA}$ shows unfairness among users for a much longer time. Weighted metrics not only show faster convergence to a high fairness level, but also give a lower Gini indicator at the end of the window length, when compared to conventional $PF^{NOMA}$.\\

\begin{figure}
\centering
\includegraphics[width=0.52\textwidth]{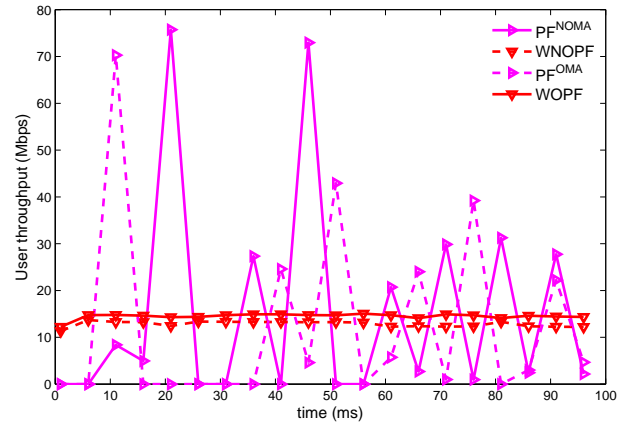}
% where an .eps filename suffix will be assumed under latex, 
% and a .pdf suffix will be assumed for pdflatex; or what has been declared
% via \DeclareGraphicsExtensions.
\caption{User throughput vs. time for NOMA-based scheduling schemes.}
\label{fig:4}
\end{figure}

In order to assess the QoE achieved by the proposed scheduling schemes, we evaluate the time required for each user to be served for the first time, referred to as the rate latency, as well the variations of its achieved rate over time. For this purpose, Fig. 7 shows the achieved rate versus time for the user experiencing the largest rate latency, for the different scheduling schemes.\\
When the conventional $PF^{NOMA}$ is used, no rate is provided for this user, for the first five scheduling slots. In addition, large rate fluctuations are observed through time. In contrast, when weighted metrics and a special treatment of the first time slot are considered, a non-zero rate is assigned for the least privileged users from the first scheduling slot, and remains stable for all the following slots. This behavior results from the fact that, at the beginning of the scheduling process (first scheduling slot), historical rates are set to zero, and $PF^{NOMA}$ uses only instantaneous achievable throughputs to choose the best candidate user set. Therefore, users experiencing bad channel conditions have a low chance to be chosen. The corresponding achieved data rates are then equal to zero. On the other side, using the proposed scheduling, the treatment of the first scheduling slot is conducted differently and users are chosen depending on their actual rates (measured during the actual scheduling period). In this case, zero rates are eliminated. Hence, latency is greatly reduced.\\
\indent For the next scheduling slots, historical rates are taken into account. For $PF^{NOMA}$, users experiencing a large $T_{k}(t)$  have less chance to be chosen, and may not be chosen at all. In this case, the use of buffering becomes mandatory and the size of the buffer should be chosen adequately to prevent overflow when peak rates occur, as a result of a high achieved throughput (high $R_{s,k}(t)$). Based on calculation, the average size of the buffer should be around 110 Mbit, for the simulation case at hand. However, in the case of the weighted proposed metrics, buffering is not needed, since only small variations between user data rates are observed, and a better QoE is achieved. Similar performance improvement is obtained for the orthogonal case in the same aforementioned conditions.\\
\begin{figure}
\centering
\includegraphics[width=0.52\textwidth]{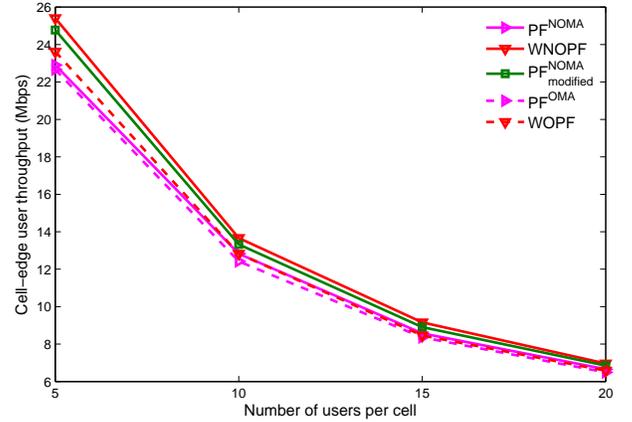}
% where an .eps filename suffix will be assumed under latex, 
% and a .pdf suffix will be assumed for pdflatex; or what has been declared
% via \DeclareGraphicsExtensions.
\caption{Cell-edge user throughput vs. number of users per cell.}
\label{fig:5}
\end{figure}
Finally, we have analyzed the effect of the proposed scheduling scheme on the cell-edge user throughput in Fig. 8. Again, the proposed weighted metrics outperform the conventional PF scheduling scheme for both access techniques, OMA and NOMA. In addition, WNOPF shows the best performance. Therefore, we can state that the incorporation of fair weights with a NOMA-based system proves to be the best combination.\\

In order to evaluate the performance of the proposed weighted metrics when premium services are considered, Tables 2 and 3 show the Gini fairness index values for two different scenarios, where three levels of services are requested: basic, silver, and gold. The number of users per group is set to 5.\\

\textbf{Scenario 1:}\\
The corresponding data rates of the three levels are set to 5 Mbps, 10 Mbps, and 15 Mbps respectively.\\
\textbf{Scenario 2:}\\
The corresponding data rates of the three levels are set to 10 Mbps, 20 Mbps, and 30 Mbps respectively.\\
In scenario 1, all users succeed in reaching their requested service data rates, and results of Table 1 show a high level of fairness achieved among users requesting the same service. However, when scenario 2 is applied, no success could be obtained but fairness is still maintained among users.\\
 
\begin{table}[!t]
%% increase table row spacing, adjust to taste
%\renewcommand{\arraystretch}{1.3}
% if using array.sty, it might be a good idea to tweak the value of
% \extrarowheight as needed to properly center the text within the cells
\caption{Gini fairness index and data rate achieved per group for Scenario 1 (100\% success)}
\label{tab:22}
\resizebox{\columnwidth}{!}{%
%\centering
%% Some packages, such as MDW tools, offer better commands for making tables
%% than the plain LaTeX2e tabular which is used here.
\begin{tabular}{|p{2.5cm}|p{2.5cm}|p{2.5cm}|}
\hline
Service & Gini fairness index & Achieved data rate per group (Mbps) \\
\hline
Basic & 0.0491 & 25.7 \\
Silver & 0.0724 & 51 \\
Gold & 0.0042 & 76.3 \\
\hline
\end{tabular}
}
\end{table}

\begin{table}[!t]
%% increase table row spacing, adjust to taste
%\renewcommand{\arraystretch}{1.3}
% if using array.sty, it might be a good idea to tweak the value of
% \extrarowheight as needed to properly center the text within the cells
\caption{Gini fairness index and data rate achieved per group for Scenario 2 (No success)}
\label{tab:2}
\resizebox{\columnwidth}{!}{%
%\centering
%% Some packages, such as MDW tools, offer better commands for making tables
%% than the plain LaTeX2e tabular which is used here.
\begin{tabular}{|p{2.5cm}|p{2.5cm}|p{2.5cm}|}
\hline
Service & Gini fairness index & Achieved data rate per group (Mbps) \\
\hline
Basic & 0.0522 & 30.2 \\
Silver & 0.0613 & 49.6\\
Gold & 0.0049 & 75.2\\
\hline
\end{tabular}
}
\end{table}

\subsection{Computational Complexity}
\indent With the aim of assessing the implementation feasibility of the different proposed schedulers, we measured the computational load of the main allocation techniques to be integrated at the BS.\\
From a complexity point of view, the proposed scheduling metric WNOPF differs from the conventional PF metric in the weight calculation. For a number of users per subband limited to 2 in NOMA, the number of candidates per subband is $\left( \begin{matrix}
   1  \\
   K  \\
\end{matrix} \right)+\left( \begin{matrix}
   2  \\
   K  \\
\end{matrix} \right)$. When listing the operations of the proposed allocation technique, we obtain that the proposed metric WNOPF increases the PF computational load by $\frac{26}{3}KS+S$ ($\simeq O(KS)$) multiplications and $-{{K}^{3}}S+\frac{3}{2}{{K}^{2}}{{S}^{2}}-\frac{4}{6}{{K}^{2}}S-\frac{3}{6}KS$ ($\simeq O(\frac{3}{2}{{K}^{2}}{{S}^{2}}-{{K}^{3}}S)$) additions.\\
\indent In order to compute the PF metric for a candidate user set containing only 1 user, $4+S$ multiplications and $1+\frac{3}{2}S$ additions are needed. For each candidate user set containing 2 multiplexed users, $13+2S$ multiplications and $6+3S$ additions are required.\\
\indent By taking account of the calculations of the terms $h^{-2\alpha}$, $h^2$, and $h^2/(N_0B/S)$ performed at the beginning of the allocation process, the classical NOMA PF requires a total of $3KS+C^1_KS(4+S)+C^2_KS(13+2S)$ multiplications which is equal to ${{K}^{2}}{{S}^{2}}+\frac{1}{2}KS+\frac{13}{2}{{K}^{2}}S$ ($\simeq O(K^2S^2)$) and $C^1_KS(1+3S/2)+C^2_KS(6+3S)$ additions which is equal to $\frac{3}{2}{{K}^{2}}{{S}^{2}}+\frac{1}{2}KS+\frac{13}{2}{{K}^{2}}S$ ($\simeq O(K^2S^2)$). Therefore, we can see that the increase in the number of multiplications in minor in comparison with that of the conventional PF, while the number of additions is almost doubled.

\section{Conclusion}
In this paper, we have proposed new weighted scheduling schemes for both NOMA and OMA multiplexing techniques. They target maximizing fairness among users, while improving the achieved capacity. Several fair weights designs have been investigated. Simulation results show that the proposed schemes allow a significant increase in the total user throughput and the long-term fairness, when compared to OMA and classic NOMA-based PF scheduler. Combining NOMA with fair weights shows the best performance. Furthermore, the proposed weighted techniques achieve a high level of fairness within each scheduling slot, which improves the QoE of each user. In addition, the proposed weighted metrics give the possibility of delivering different levels of QoS which can be very useful for certain applications. The study conducted here with two scheduled users per subband can be easily adapted to a larger number of users. We are currently undergoing further research to reduce the complexity of the PF scheduler by introducing an iterative allocation scheme, and also to study the applicability of our framework in the context of uplink transmission.

% or


%\appendices

%\section{}

%\section*{Acknowledgment}


\ifCLASSOPTIONcaptionsoff
  \newpage
\fi

\end{document}